\documentclass[12pt,preprint]{aastex}

\shorttitle{A new satellite in Ursa Minor}
\shortauthors{Mu\~noz et al.}

\begin{document}

\title{The Discovery of an Ultra-Faint Star Cluster in the Constellation of Ursa Minor\altaffilmark{1}}

\author{
R.\ R.\ Mu\~noz\altaffilmark{2,3},
M.\ Geha\altaffilmark{2},
P.\ C\^ot\'e\altaffilmark{4},
L.\ C.\ Vargas\altaffilmark{2},
F.\ A.\ Santana\altaffilmark{3},
P.\ Stetson\altaffilmark{4},
J.\ D.\ Simon\altaffilmark{5} \&
S.\ G.\ Djorgovski\altaffilmark{6,7}
}

\altaffiltext{1}{Based on observations obtained at the Canada-France-Hawaii Telescope (CFHT) 
which is operated by the National Research Council of Canada, the Institut National des 
Sciences de l'Univers of the Centre National de la Recherche Scientifique of France,  and 
the University of Hawaii.  Spectroscopic data presented herein were obtained at the 
W.~M.~Keck Observatory, which is operated as a scientific partnership among the California 
Institute of Technology, the University of California and the National Aeronautics and 
Space Administration.}

\altaffiltext{2}{Astronomy Department, Yale University, New Haven, CT 06520, USA}

\altaffiltext{3}{Departamento de Astronom\'ia, Universidad de Chile, Camino El 
Observatorio 1515, Las Condes, Santiago, Chile (rmunoz@das.uchile.cl)}

\altaffiltext{4}{Herzberg Institute of Astrophysics, National Research Council of Canada, 
Victoria, BC, V9E 2E7, Canada}

\altaffiltext{5}{Observatories of the Carnegie Institution of Washington, 813 Santa 
Barbara St., Pasadena, CA 91101, USA}

\altaffiltext{6}{Astronomy Department, California Institute of Technology, 
Pasadena, CA, 91125, USA}

\altaffiltext{7}{Distinguished Visiting Professor, King Abdulaziz University, Jeddah,
Saudi Arabia.}

\begin{abstract}

  We report the discovery of a new ultra-faint globular cluster in the
  constellation of Ursa Minor, based on stellar photometry from the
  MegaCam imager at the Canada-France-Hawaii Telescope (CFHT). We find
  that this cluster, Mu\~noz~1, is located at a distance of
  $45\pm5$\,kpc and at a projected distance of only $45\arcmin$ from
  the center of the Ursa Minor dSph galaxy.  Using a Maximum
  Likelihood technique we measure a half-light radius of $0\arcmin.5$,
  or equivalently $7$\,pc and an ellipticity consistent with being
  zero.  We estimate its absolute magnitude to be $M_{V}=-0.4\pm0.9$,
  which corresponds to $L_{V}=120^{+160}_{-65}$\,L$_{\sun}$ and  we
  measure a heliocentric radial velocity of $-137\pm 4$ km s$^{-1}$ based on
  Keck/DEIMOS spectroscopy.  This new satellite is separate from Ursa Minor by
  $\sim30$\,kpc and 110 km s$^{-1}$ suggesting the cluster is not obviously
  associated with the dSph, despite the very close angular
  separation.  Based on its photometric properties and structural
  parameters we conclude that Mu\~noz~1 is a new ultra-faint stellar
  cluster. Along with Segue~3 this is one of the faintest stellar
  clusters known to date.

\end{abstract}


\section{Introduction}

Over the last seven years, and thanks to the advent of the
Sloan Digital Sky Survey \citep[SDSS;][]{york00a}, seventeen new
Milky Way satellites have been discovered 
\citep[e.g.,][]{willman05a,zucker06a,zucker06b,belokurov06a,belokurov06b,irwin07a}.
The majority of these objects correspond to a new class of
ultra low-luminosity dwarf galaxies, based on their kinematic
and metallicity properties \citep[e.g.,][]{munoz06a,martin07a,simon07a},
with luminosities ranging from $M_{V}\sim-8$ for the brighter of
these objects down to an extreme $M_{V}\sim-1.5$ for the faintest \citep{martin08a}.
Given their extremely low stellar contents, 
they are commonly referred to as the Ultra-Faint Dwarfs (UFDs). A few other
systems are clearly low luminosity globular clusters, which include
Koposov~1 and 2 ($M_{V}=-1$ and $-2$ respectively; \citealt{koposov07a}) 
and the extreme case of Segue~3 ($M_{V}\sim0.0$; \citealt{belokurov10a,fadely11a}).

Despite the enormous success of satellites searches in SDSS, this survey 
covers only about a fourth of the sky down to a magnitude limit of $r\sim22.5$.
In practice this means that the faintest objects can only be detected out
to distances smaller than a few tens of kiloparsecs (see Fig. 9 in \citealt{tollerud08a}),
and thus the majority of the Milky Way's virial volume remains
unsearched for UFDs 
and other faint stellar systems.

In this Letter we present the discovery of a new extremely
low-luminosity, outer halo stellar cluster,
in the outskirts of the Ursa Minor dwarf spheroidal (dSph) galaxy.
This discovery was made just outside the SDSS footprint
using significantly deeper photometry which allowed us
to probe farther into the Milky Way halo.
In \S2 we detail the data-set and the discovery. In \S3
we present the results from the photometric analysis and in \S4 we 
show 
radial velocity data recently obtained. Finally, in
\S5 we discuss and summarize our results.
 
\section{Data and Discovery}

The discovery of this new system was made serendipitously
while analyzing photometric data for the Ursa Minor dSph galaxy
taken with the MegaCam imager at the Canada-France-Hawaii Telescope (CFHT).
MegaCam is a wide-field imager consisting of 36 $2048\times4612$ pixel CCDs,
covering almost a full $1\times1$\,deg$^{2}$ field of view with a
pixel scale of $0".187$\,pixel$^{-1}$.
These data were taken as part of a larger program aimed at
obtaining deep wide-field imaging of all bound stellar
over-densities in the Milky Way halo beyond 25\,kpc 
(R. R. Mu\~noz et al. 2012, in preparation).
Ursa Minor was observed on the nights of UT June 12-14 and July 07-13, 2010
under dark conditions with typical seeing of $0.7-0".9$.
Four different, slightly overlapping fields were observed for a total area
coverage of nearly $2\times2$\,deg$^{2}$. In each field, the center
of the dSph was placed in one of the corners so that, when combined, the
galaxy is located at the center of the covered area.
We obtained six dithered exposures of $360$ seconds in both
the $g-$ and $r-$band. A standard dithering pattern was used to cover
both small and large gaps present between the chips.
    
MegaCam images are delivered to the user pre-processed by the CFHT team
using the ``Elixir" package \citep{magnier04a}.
Subsequent photometric measurements were carried out using first
DAOPHOT/Allstar, and later running the ALLFRAME package on the processed
frames, following the procedure outlined in \citet{stetson94a}.
Finally, astrometric solutions were calculated using the freely
available SCAMP\footnote{See http:/www.astromatic.net/software/scamp.} code, and 
photometric calibration was carried out by direct comparison with data from the 
SDSS-Data Release 7 (DR7; \citealt{abazajian09a}).
For more details on the method see \citet{munoz10a}.

\begin{figure}
\epsscale{0.4}
\plotone{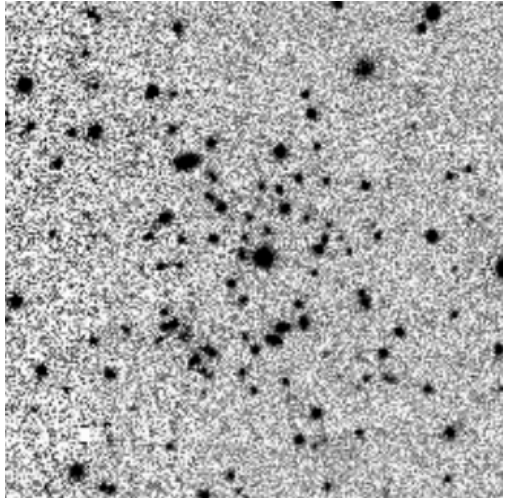}
\caption{$1\arcmin.2 \times 1\arcmin.2$,  $r-$band view of Mu\~noz~1. The image is roughly
centered on the cluster. North is up, east is to the left.
}  \label{picture1}
\end{figure}

\begin{figure}
\epsscale{1.0}
\plotone{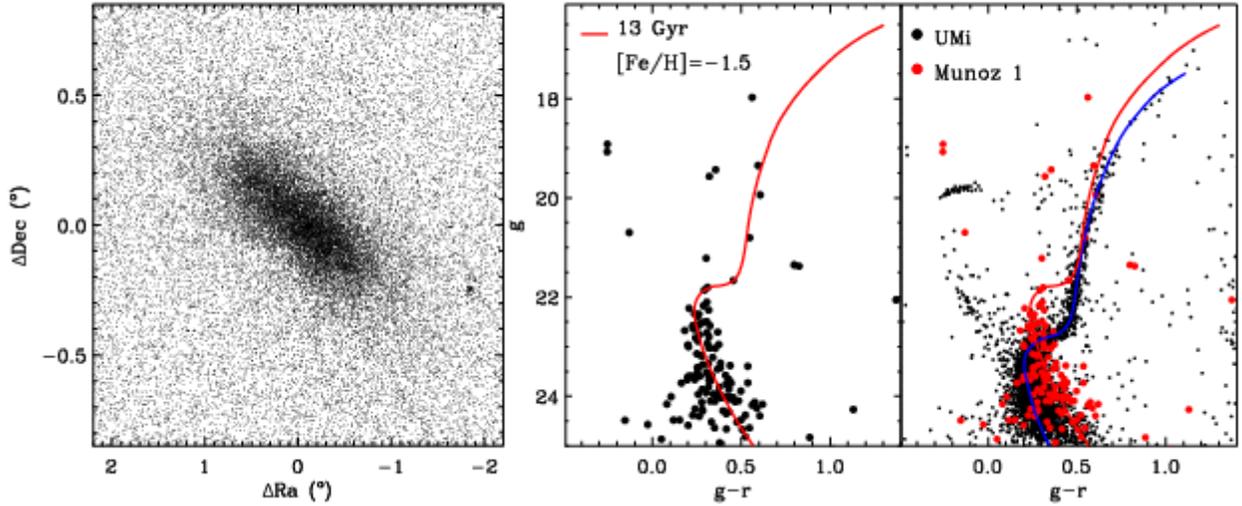}
\caption{
{\it left:} Star count map of the Ursa Minor dSph field. The primary object in this field
is the dSph galaxy. Mu\~noz~1 lies $45\arcmin$ away from the center of the field
in the south-west direction, near ($-1.8^{\circ}$,$-0.25^{\circ}$).
{\it middle:} $g$ vs $g-r$ CMD of stars in the Mu\~noz~1 region 
(within $2\arcmin$ of the measured center). 
The best fit isochrone is overplotted, corresponding to a 
Dartmouth \citep{dotter08a}, 12.5\,Gyr, [Fe/H]$=-1.5$ at a distance
 of $\sim45$\,kpc using a reddening value of $E(g-r)=0.027$, 
adopted from 
the \citet{schlegel98a} maps. 
{\it right:} CMD of Ursa Minor stars (small black points) 
overplotted on Mu\~noz~1 stars (large red symbols). Red and blue solid lines represent 
best fit isochrones for both the Ursa Minor dSph and Mu\~noz~1 respectively.
    }  \label{munoz1_cmd}
\end{figure}

As seen in Figure~\ref{picture1} and the left panel of Figure~\ref{munoz1_cmd}, 
we discovered a centrally concentrated
over-density of stars $45\arcmin$ to the south-west of Ursa Minor.
Based on the structural parameters determined below, we presume
this object to be an ultra-faint globular cluster and therefore name
it Mu\~noz~1. In \S5 we discuss in more depth our reasoning for this classification.
Figure~\ref{munoz1_cmd} also shows the color-magnitude diagram (CMD) of the central
region of Mu\~noz~1 where a distinct main-sequence turn-off is observed. 
Assuming the stars in the red-giant branch (RGB) region are members of
the cluster we fit an isochrone at a distance of 45\,kpc, an age of 12.5\,Gyr 
and [Fe/H]$=-1.5$. We note that this fit is very tentative given the very
low number of stars that belong to the system.
The right panel
of the figure compares the CMDs of UMi and the new object showing a clear
difference in distance modulus.

\section{Structural Properties}

We carry out a maximum-likelihood analysis of the 
photometric data for Mu\~noz~1 following the method
of \citet{martin08a} as described in \citet{munoz10a}.
This method starts by assuming a shape for the underlying light
distribution; in this case we have tried three of the 
most commonly used profiles for UFDs, a King \citep{king62a},
exponential and Plummer \citep{plummer11a} density laws. 
We then fit simultaneously the structural parameters of the
object, i.e., scale length (half-light radius or King core and tidal
radii depending on the profile), coordinates of the center
of the cluster, ellipticity, position angle and background
density.  To estimate parameter uncertainties we carry
out a bootstrap analysis of $10000$ realizations of the
photometric data.

Table~1 shows the resulting parameters.
We obtain a half-light radius of $7.1\pm2.1$\,pc using both
an exponential and Plummer profile, whereas a King profile
yields $r_{\rm core}=4.5\pm1.9$\,pc and $r_{\rm tidal}=65\pm28$\,pc.

\begin{figure}
\epsscale{1.0}
\plotone{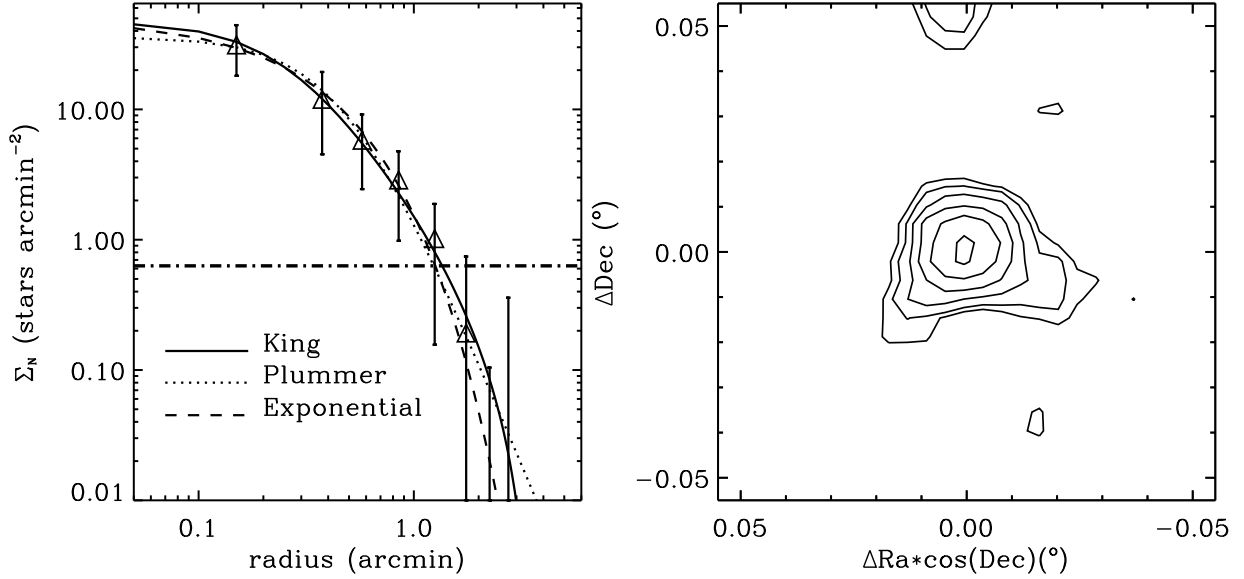}
\caption{{\it Left:} Number density profile for Mu\~noz~1. Plummer, King and exponential
profiles have been fitted to the data using a maximum likelihood method. The dot-dashed line
represents the measured background density. {\it Right:} Iso-density contours
for Mu\~noz~1. Contours level correspond to 2,3,5,8,12 and 18$- \sigma$ level over
the background density.
}  \label{profile}
\end{figure}

Figure~\ref{profile} shows the number density profile for Mu\~noz~1. 
We have overplotted the three best-fit profiles,
all of which give a reasonable description of
the light distribution. The right panel of this figure shows
a density contour map of the object, where the contours
represent 2, 3, 5, 8, 12 and $18- \sigma$ levels above the
background density. These contours show no evident elongation
or tidal features. Applying a bootstrap analysis where we generate
$10000$ realizations of the photometric data \citep{walsh08a,munoz10a},
we determine that
the asymmetries observed in the outer parts,
toward the south and west directions, are not statistically significant
due to the very low number of stars.

In addition to the structural parameters, we estimate the
absolute magnitude of the new object. We follow the method described
in \citet{munoz10a} which relies in the total number
of stars that belong to the cluster and not on the sum of their
fluxes. As shown by \citet{martin08a} the low number of stars
identified in the ultra-faint systems, especially true in the case of
Mu\~noz~1, makes traditional methods of adding the individual
fluxes of the member stars very sensitive to the inclusion (or
exclusion) of potential members, especially at brighter magnitudes.
To alleviate shot noise issues we use an alternative
method based on a model stellar population. For this we use
a theoretical luminosity function that best describes the
photometric properties of the cluster, in this case
a $12.5$\,Gyr old population with [Fe/H]$=-1.5$ \citep{dotter08a}.
We then integrate the LF to obtain the total flux down
to a given magnitude limit. The last step is to scale this flux
using the total number of stars that belong to the cluster down
to the same magnitude limit. 
We estimate the corresponding 
uncertainty 
through a bootstrap analysis. 
This yields $M_{V}=-0.4\pm0.9$\,mag
assuming a Chabrier initial mass function \citep{chabrier01a}, which
corresponds to $L_{V}=120^{+160}_{-65}$\,L$_{\sun}$. Only
one old star cluster has a published luminosity lower than
Mu\~noz~1, the Segue~3 stellar cluster with a total
luminosity of $M_{V}=-0.0\pm0.8$\,mag (Fadely et al. 2011),
although given the large uncertainties in both measurements
it is unclear which cluster is actually fainter.

\section{Spectroscopic Properties}

Spectroscopic data of photometrically selected Mu\~noz~1 candidate stars
were taken on UT May 28, 2011 with the Keck II 10-m
telescope and the DEIMOS spectrograph \citep{Faber03a}.  One
Keck/DEIMOS multi-slit mask was observed with the
1200~line~mm$^{-1}$\,grating covering a wavelength region
$6400-9100\mbox{\AA}$ with a spectral dispersion of $0.33\mbox{\AA}$.
The mask was observed for 7800\,seconds under relatively poor seeing
conditions which varied between $1.5-2''$.  Spectra were reduced
using a modified version of the spec2d software pipeline (version
1.1.4) developed by the DEEP2 team, and radial velocities were determined
using the software described by \citet{simon07a}.  For additional
details, we refer the reader to \citet{geha09a}.

Radial velocities were successfully measured for 24 of the 47
extracted spectra.  Stars for which we could not measure a velocity
were primarily very low $S/N$ spectra of faint objects near the expected
main sequence turn-off of Mu\~noz~1 ($r\sim 22-22.5$).  

\begin{figure}
\epsscale{1.0}
\plotone{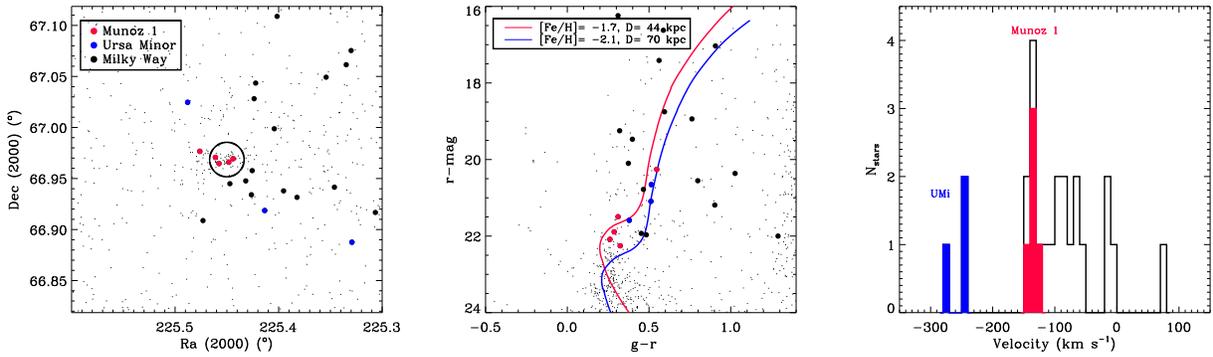}
\caption{Keck/DEIMOS spectroscopic coverage of Mu\~noz~1, showing the
  spatial distribution ({\it left}), color-magnitude diagram ({\it
    middle}) and velocity distribution of targets ({\it right}).  Red
  symbols in each panel indicate stars possibly associated with Mu\~noz~1,
  while blue symbols are stars likely associated with Ursa
  Minor.}  \label{umi_spec}
\end{figure}

The resulting velocity distribution is shown in
Figure~\ref{umi_spec}, along with the corresponding spatial and
color-magnitude distributions.
While the velocity histogram does not show a clear significant
peak that could be interpreted as the velocity signature of Mu\~noz~1, if
we restrict ourselves to the most centrally concentrated stars, what appears as a
cold peak is indeed discerned.
Using the four stars closest to the derived center of the cluster, plus
a fifth star possibly associated with it based on its radial velocity,
distance to the cluster and position in the CMD, we estimate
a mean velocity of $-137\pm4$\,km s$^{-1}$.
Given the small number of stars and large velocity uncertainties, we estimate
only a one sigma upper limit on the velocity dispersion of 
$\sigma < 4.7$\,km s$^{-1}$.   
Given the size and luminosity of Mu\~noz~1, we predict its
velocity dispersion to be between $0.2-0.3$\,km s$^{-1}$ based on
\citet{wolf10a} and a mass-to-light ratio consistent with old stars
only.   Our measured velocity dispersion is consistent with this
value, but we cannot rule out a scenario where the mass of Mu\~noz~1 is
significantly larger than its stellar mass.

Two other stars with radial velocities in the vicinity of this peak 
do not live near the CMD of Mu\~noz~1, nor are they
close to the center of the object, and therefore we consider them
less likely to be associated with the cluster.
We also observe three stars consistent with the published radial
velocity of Ursa Minor of $-245$ km s$^{-1}$ \citep{munoz05a}.  These stars are
less spatially concentrated than the Mu\~noz~1 stars and yield a mean
velocity of $-253\pm6$\,km s$^{-1}$, in good agreement with the observed value 
for the dSph galaxy.

A subset of stars in our 
Keck/DEIMOS kinematic sample 
have sufficient
$S/N$ (at least $10\mbox{\AA}^{-1}$) to determine their metallicities.
We determine [Fe/H] based on fitting synthetic spectra, as
described in \citet{kirby08a}.
This medium-resolution technique allows us to calculate $T_{\rm eff}$ and [Fe/H] by
matching an observed Keck/DEIMOS spectrum to a grid of models with 
different $T_{\rm eff}$ and [Fe/H]. 
This is in contrast to most high resolution abundances, which are
based on equivalent widths and require resolved, unblended
lines. Our spectral resolution is not sufficient for such type of analysis. 
We note, however, that our well-characterized noise array enables
us to calculate realistic error bars. 

We measure metallicities for two stars kinematically associated with Ursa Minor and one star
associated with Mu\~noz~1. The two Ursa Minor stars have metallicities
[Fe/H]$=-2.78\pm0.60$ and $-2.30\pm0.68$, while the Mu\~noz~1 star has 
[Fe/H]$=-1.46\pm0.32$. These results are
consistent with the isochrone-derived metallicity.
Due to the paucity of Fe~II absorption lines in cool RGB spectra, we are unable
to calculate surface gravity from the spectroscopic data. Hence, to estimate [Fe/H] we
had to fix log\,$g$ from the best-fit isochrone. 
The code also outputs a value for $T_{\rm eff}$, which can be compared to the 
value obtained using the 
de-reddened CFHT photometry.
For the case of the cluster star, we obtain a 
$T_{\rm eff} = 5160\pm320$\,K, in good agreement with the
photometric value of $T_{\rm eff} = 5290$\,K.

\section{Summary and Discussion}

We report the discovery of a new ultra-faint stellar system, 
Mu\~noz~1, in the
Ursa Minor constellation, located only 45$\arcmin$ from the center of the
Ursa Minor dSph.
The discovery was made serendipitously while analyzing photometric data for 
Ursa Minor, taken 
with the wide-field MegaCam imager on the CFHT.
We visually fit a 12.5\,Gyr isochrone
from the Dartmouth database \citep{dotter08a},
with a metallicity of [Fe/H]$=-1.5$ at a distance
 of $\sim45$\,kpc. While the close angular distance between these two Galactic 
satellites might suggest an association, we deem it unlikely.
We have estimated the escape velocity from Ursa Minor at the projected
distance of Mu\~noz~1, i.e., 1\,kpc, assuming the dSph is a point mass of
$1\times10^{8}$\,M$_{\sun}$ \citep{wolf10a}. We obtain $v_{\rm esc}\sim30$ km s$^{-1}$, much
smaller than the $\sim110$\,km s$^{-1}$ of radial velocity difference.
Moreover, at 30\,kpc, the distance separating the cluster from the dSph along the 
line of sight, $v_{\rm esc}$ decreases down to only a few km s$^{-1}$.

Using a maximum-likelihood technique, following the method of \citet{martin08a}, 
we have simultaneously derived structural parameters along with 
background density for Mu\~noz~1. 
Assuming different density profiles we have derived consistent half light radii of
 $r_{h,exp}=7.1\pm2.8$\,pc, $r_{h,P}=7.1\pm2.1$\,pc for exponential and Plummer 
profiles respectively, 
and a King core and tidal radii of $r_c=4.5\pm1.9$ and $r_t=65\pm28$\,pc
respectively.
The total luminosity of Mu\~noz~1, on the other hand, was derived using the method 
described in \citet{munoz10a}, which relies in the total number of stars in the object 
instead of their individual fluxes, obtaining an absolute magnitude of
$M_{V}=-0.4\pm0.9$, equivalent to $L_{V}=120_{-65}^{+160}L_{\sun}$.
Based on its photometric properties, we argue that this new Galactic satellite 
corresponds to an extremely faint globular cluster. 
Since the discovery of the UFDs, there has been some debate in the literature
about the classification of stellar over-densities based solely on 
photometric information. It has been argued, for instance, that objects
like Segue~1 and Coma Berenices should be considered large globular clusters
based on the assumption that galaxies cannot have scale lengths smaller than 
$\sim100$\,pc \citep{gilmore07a}. On the other hand, spectroscopic information, i.e., 
velocity dispersions
and metallicities, have been used to argue that despite 
their sizes, these systems correspond to dark matter dominated
galaxies \citep{simon07a,geha09a}. 
For Mu\~noz~1, the spectroscopic information is insufficient to determine
unambiguously whether the object is dark matter dominated or not. However, 
this satellite is significantly smaller than any of the known dwarf galaxies, 
including the UFDs, it lies within the size distribution of other Milky Way
globular clusters with structural parameters similar to other outer halo clusters
like Palomar~13 \citep{bradford11a}, and while in principle it could be the smallest 
known galaxy, without a reliable measurement of its velocity dispersion of metallicity spread,
we view it as much more likely that it is a globular cluster.
The luminosity derived for this object, makes it one of the faintest globular cluster 
discovered, only nominally surpassed by Segue~3, which has an absolute magnitude 
of $M_{V}=0.0\pm0.8$ \citep{fadely11a}. Given the large uncertainties of both 
measurements, it is unclear which one of them is indeed fainter.
The density profile of this object is well fitted by typical
cluster density profiles showing no obvious
hints for 
significant tidal disruption.

The discovery of this cluster was made in the context of a larger
program aimed at obtaining deep and wide-field photometry for all
outer halo satellites regardless of morphological classification.
At the time of this
publication, we have imaged roughly 40\,deg$^{2}$ down to a magnitude
limit of $r\sim24.5-25$ depending on the object.
As pointed out by \citet{tollerud08a}, surveys deeper than SDSS are
bound to uncover fainter and more distant ultra-faint systems.
Naively extrapolating the discovery of at least one ultra-faint cluster
in our surveyed area to the entire sky, yields potentially
a thousand similar objects still uncovered.
While this number is not intended as a firm prediction given its
large uncertainty, it foreshadows a much more complex picture of the globular 
cluster system in the outer halo than we have today.

We thank the referee for useful comments that helped improve this paper.
We acknowledge Robert Zinn for useful discussions.  This work was
supported in part by the facilities and staff of the Yale University
Faculty of Arts and Sciences High Performance Computing Center.
R.R.M.~acknowledges support from the GEMINI-CONICYT Fund, allocated to the project
N$^{\circ}32080010$, from CONICYT through project BASAL PFB-06 and the Fondo Nacional
de Investigaci\'on Cient\'ifica y Tecnol\'ogica (Fondecyt project N$^{\circ}1120013$).
M.~G.~acknowledges support from the National Science Foundation under
award number AST-0908752 and the Alfred P.~Sloan Foundation. S.G.D.~acknowledges
partial support from the NSF grant AST-0909182.

\begin{deluxetable}{lcc}
\tablewidth{0pt.}
\tablecaption{Parameters of Mu\~noz~1}
\tablehead{
\colhead{Parameter} & \colhead{Mean} & \colhead{Uncertainty}\\
}
\startdata
$\alpha_{0,exp}$ (h~m~s)&15:01:48.02&$\pm13''$ \\
$\delta_{0,exp}$ (d~m~s)&+66:58:07.3&$\pm8''$ \\
$r_{h,exp}$ (arcmin) &0.49&$\pm0.19$ \\
$r_{h,exp}$ (pc) & 7.1 & $\pm2.8$ \\
$r_{h,\rm P}$ (arcmin) &0.49&$\pm0.15$ \\
$r_{h,\rm P}$ (pc) &7.1&$\pm2.1$\\
\hline
$r_c$ (arcmin)&15"&$\pm9"$  \\
$r_c$ (pc)&4.5&$\pm1.9$ \\
$r_t$ (arcmin)&3.6&$\pm2.2$ \\
$r_t$ (pc)&65&$\pm28$ \\
\hline
$M_{V}$ (Chabrier)& $-0.4$ & $\pm0.9$ \\
$L_{V}$  (L$_{\odot}$)& $120$ & $+160,-65$ \\
$\mu_{0,V}$ (mag arcsec$^{-2}$) & $26.3$ & $+1.6,-2.1$ \\
\hline
$v_{r}$ (km s$^{-1}$)& $-137$ & $\pm4$ \\
$d$ (kpc)& $45$ & $\pm5$ 
\enddata 
\end{deluxetable}

\end{document}